\documentclass[11pt]{article}

\usepackage{graphicx, verbatim}
\usepackage{amssymb}
\usepackage{alltt}
\usepackage{amsmath, amssymb, amsfonts, amscd, xspace, pifont, amsthm}
\usepackage{mathrsfs}
\usepackage{algorithm}

\usepackage{authblk}
\usepackage{xcolor}
\usepackage[normalem]{ulem} 
\usepackage{soul}

\newcommand{\V}[1]{\ensuremath{\boldsymbol{#1}}\xspace}

\def\threeImages#1#2#3#4#5#6#7#8#9
{
	\centerline{\hfill\makebox[#2]{#3}\hfill\makebox[#5]{#6}\hfill\makebox[#8]{#9}\hfill}
	\centerline{\hfill
		\includegraphics[width=#2]{#1}
		\hfill
		\includegraphics[width=#5]{#4}
		\hfill
		\includegraphics[width=#8]{#7}
		\hfill}
}

\def\twoImages#1#2#3#4#5#6
{
	\centerline{\hfill\makebox[#2]{#3}\hfill\makebox[#5]{#6}\hfill}
	\centerline{\hfill
		\includegraphics[width=#2]{#1}
		\hfill
		\includegraphics[width=#5]{#4}
		\hfill}
}

\newtheorem{thm}{Theorem}

\newtheorem{remark}{Remark}

\newtheorem{definition}{Definition}

\DeclareMathOperator*{\argmax}{arg\,max}

\begin{document}
	\title{A Survey on Theoretical Advances of Community Detection in Networks}

	\author{Yunpeng Zhao\thanks{Department of Statistics, George Mason University. The author gratefully acknowledges NSF DMS 1513004}}
	
	\date{}

		\maketitle
	
	\begin{abstract}
		Real-world networks usually have community structure, that is, nodes are grouped into densely connected communities. Community detection is one of the most popular and best-studied research topics in network science and has attracted attention in many different fields, including computer science, statistics, social sciences, among others. Numerous approaches for community detection have been proposed in literature, from ad-hoc algorithms to systematic model-based approaches. The large number of available methods leads to a fundamental question: whether a certain method can provide consistent estimates of community labels. The stochastic blockmodel (SBM) and its variants provide a convenient framework for the study of such problems. This article is a survey on the recent theoretical advances of community detection. The authors review a number of community detection methods and their theoretical properties, including graph cut methods, profile likelihoods, the pseudo-likelihood method, the variational method, belief propagation, spectral clustering, and semidefinite relaxations of the SBM. The authors also briefly discuss other research topics in community detection such as robust community detection, community detection with nodal covariates and model selection, as well as suggest a few possible directions for future research. 
	\end{abstract}

	\section*{\sffamily \Large INTRODUCTION} 
	
	Network science is the study of networks (or graphs) as a representation of relations (called \textit{edges} or \textit{links}) between objects (called \textit{vertices} or \textit{nodes}) \cite{Newman2010,Kolaczyk2009}. Networks have become one of the most common data structure. One famous example is the Internet, which is the physical network, composed of computers, routers and modems linked by electronic, optical and wireless networking technologies. Other well-known examples include online social networks such as \textit{Facebook} and \textit{LinkedIn}, citation networks, gene regulatory networks, protein-protein interaction networks, food webs, among others. In the past decades, network science has drawn a lot of attention in many different branches of science and engineering, for example, computer science \cite{Getoor2005}, physics \cite{Albert&Barabasi2002,Newman2003Review}, biology \cite{Karlebach08}, social sciences \cite{Wasserman1994,Liben2007} and economics \cite{Jackson2008}. It is worth mentioning that network analysis has also become an active research area in statistics. A number of probabilistic and statistical models have been proposed. Typical examples include the Erd{\"o}s-R\'enyi random graph model \cite{Erdos59}, exponential random graph models \cite{Frank&Strauss1986,Wasserman1996}, latent space models \cite{Hoff2002}, stochastic blockmodels \cite{Holland83}, the preferential attachment model \cite{Barabasi&Albert1999}, among others 
	(see Goldenberg et al. \cite{Goldenberg2010} for a comprehensive review). 
	
	Most networks have community structure, that is, nodes are grouped into densely connected \textit{communities} or \textit{clusters}. Detection of such communities is one of the most popular research topics in network science. The precise definition of community is difficult to formalize, and even no full agreement is reached on the general notion of community by researchers in different fields. We will offer some discussion on this point after introducing stochastic blockmodels in the next section. Readers can also see Fortunato and Hric \cite{Fortunato16} for more discussion. In this article, we adopt the most commonly used concept of community, that is, a community is a group of nodes with many links between themselves and fewer links to the rest of the network. Correspondingly, the goal of community detection is to partition the node set into overlapping or non-overlapping cohesive communities. We focus on non-overlapping community detection in this article. 
	
	Classical community detection methods in the literature can be loosely classified into three categories. Methods in the first category are algorithm-based, such as hierarchical clustering, in which nodes progressively agglomerate into communities according to a certain similarity measure of nodes, and edge removal, in which edges
	are progressively removed until disconnected components appear (see Newman \cite{Newman2004Review} for a more comprehensive review of algorithm-based approaches). The second category consists of criterion-based methods, which optimizes some criteria over all possible network partitions. Examples of these criteria include the ratio cut \cite{Wei&Cheng1989}, the normalized cut \cite{Shi&Malik2000}, and Newman-Girvan modularity \cite{NewmanPNAS} (see review papers \cite{Fortunato16,Fortunato2010} for more details). Methods in the third category are model-based. Such methods rely on fitting a probabilistic model for a network with community structure, in which the community labels are latent and to be identified. The best studied model for community detection is the stochastic blockmodel (SBM) \cite{Holland83,Snijders&Nowicki1997,Nowicki2001}, which plays a central role in the theoretical analysis of community detection. Other examples include the degree-corrected stochastic blockmodel \cite{Karrer10,Zhaoetal2012}, the mixed membership stochastic blockmodel \cite{Airoldi2008}, the latent position cluster model \cite{Handcock2007}, etc. It is worth adding two comments here before proceeding. Firstly, there is no clear distinction among these categories. For instance, fitting a probabilistic model usually leads to a criterion to be optimized and the optimization eventually relies on an algorithm. Secondly, community detection in networks is an analogy of cluster analysis in multivariate data. Some community detection methods are borrowed from classical cluster analysis. For instance, hierarchical clustering in community detection is essentially identical to the algorithm in cluster analysis. The only difference is the definition of similarity measures, that is, similarity measures used in community detection are usually based on network topology while similarity measures in clustering are based on distances between data points. From the algorithmic point of view, the normalized cut is also identical to the corresponding algorithm in image segmentation \cite{Shi&Malik2000}, and has become even more straightforward in community detection. That is, the normalized cut in image segmentation requires the construction of a similarity matrix from image data while the algorithm can directly use the adjacency matrix of a network as the input. The definition of adjacency matrix will be given in the next section. 
	
	A fundamental question of community detection is whether a proposed method is able to correctly identify the community labels in principle. Or more precisely in statistical terminology, a fundamental theoretical question is whether a certain method can provide consistent estimates of community labels. Despite the conceptual similarity, community detection in networks is fundamentally different from clustering in multivariate data from a theoretical point of view. The structure of network data is unique. Unlike multivariate data, which are typically assumed to be independently and identically distributed, a network is represented by a single adjacency matrix, and thereby no \textit{replicates} in the usual sense are available. This unique data structure offers a great challenge in theoretical studies of community detection. 
	
	The SBM provides a natural framework for theoretical analysis of community detection. Under the SBM, many existing community detection methods are better understood, and numerous new methods have been proposed and analyzed. These are the main focus of the current review article. The rest of this article is organized as follows. After introducing basic notations, we give the precise definition of the SBM. Next, we introduce some first results on consistency of community detection under the SBM and its variants. These results study the global optimizers of certain detection criteria over all possible label assignments. However, the global optimization of these criteria are in principle NP hard. Therefore, many computationally feasible methods have been proposed. Mainstream approaches include the pseudo-likelihood method, the variational method, belief propagation, spectral clustering, and semidefinite programming for the SBM. Many of these methods have been theoretically justified under the SBM and the corresponding results will be discussed in this article. In the last section, we will briefly discuss other research topics in community detection, including robust community detection, community detection with nodal covariates and model selection, as well as suggest a few possible directions for future research. 
	
	\section*{\sffamily \Large STOCHASTIC BLOCKMODELS} 
	
	We begin by introducing basic notation. A network or a graph can be denoted by an ordered pair $N=(V,E)$, where $V$ is the set of nodes and $E$ is the set of edges. Without any loss of generality, we will assume $V=\{1,...,n\}$. A network with size $n$ can be represented by an $n\times n$ adjacency matrix $A=[A_{ij}]$, where 
	\[
	A_{ij}=\begin{cases}
	1 & $if there is an edge between $ i $ and $ j, \\
	0 & $otherwise$. \\
	\end{cases}
	\]
	Unless otherwise specified, we consider unweighted and undirected networks, and thus $A$ is a binary symmetric matrix. And we assume that there is no self-loop in the network, i.e., $A_{ii}=0$, for $i=1,...,n$. 
	
	We now formulate community detection and give the definition of the stochastic blockmodel (SBM)  \cite{Bickel&Chen2009,Zhaoetal2012,Amini.et.al.2013}. The goal of community detection is to find a disjoint partition $V = V_1 \cup \dots \cup V_K$, or equivalently node labels $\V{e}=\{e_1,...,e_n\}$, where $e_i$ is the label of node $i$ and takes values in $\{1,2,...,K\}$. The SBM is perhaps the most commonly used model for representing a network with community structure. Under the SBM, a network is generated in two steps:
	\begin{itemize}
		\item [1)] The true node labels $\V{c}=\{c_1,...,c_n\}$ are drawn independently from $Multinomial(1,\pi)$, where $\pi=(\pi_1,...,\pi_K)$.
		\item [2)] Given the labels $\V{c}$, the edge variables $A_{ij}$ for $i<j$ are independent Bernoulli variables with 
		\begin{align}\label{SBM_prob}
		\mathbb{E}[A_{ij}|\V{c}]=P_{c_ic_j},
		\end{align}
		where $P=[P_{ab}]$ is a $K \times K$ symmetric matrix. 
	\end{itemize}
	Before we proceed to discuss detection methods and theoretical results under the SBM, it is worth adding several remarks on the model itself. 
	
	Firstly, the SBM can be understood as an analogy the Gaussian mixture model, for readers familiar with model-based clustering in multivariate analysis \cite{Fraley02}. But there is a crucial difference: The link probability for $A_{ij}$ under the SBM depends on two community labels $c_i$ and $c_j$, unlike the Gaussian mixture model. In the author's opinion, this ``two-dimensional'' structure is the root cause of many theoretical and computational challenges. 
	
	Secondly, under the SBM, two nodes within a group are \textit{stochastically equivalent} in terms of their link probabilities to other nodes \cite{Holland83}. Or intuitively speaking, two nodes within the same group play a similar role in the network. This leads back to the question of what is a community. As mentioned in the introduction, we treat community as a group of nodes with many links between themselves and fewer links to the rest of the network throughout this paper. But one can also define community as a group of nodes with similar statistical behavior. And to the best of our knowledge, historically the SBM was introduced by social scientists to in order to model the latter case. In order to model communities in the usual sense, the SBM needs constraint on parameters that the within-group densities are larger than the cross-group densities, although many theoretical results do not require this constraint.

	Thirdly, the community labels $\V{c}$ were treated as either random or deterministic in different literatures for their own technical conveniences. But practically it makes little difference since $\V{c}$ is unknown in either case (either \textit{latent} random variables or \textit{unknown} fixed parameters). 
	
	Fourthly, note that the edge variables $A_{ij}$ are independent given the labels, and with $c_i=k$ and $c_j=l$, $A_{ij}$ are identically distributed. Therefore, the SBM essentially assumes edge variables to be independently and identically distributed. This makes the SBM a convenient working model for studying asymptotic properties of community detection as the size of the network goes into infinity. That being said, these asymptotic results are still highly nontrivial even under the assumptions of the SBM, because the number of community labels to be identified also grows with the network size.

	\section*{\sffamily \Large FIRST RESULTS ON CONSISTENCY OF COMMUNITY DETECTION} 
	
	In this section, we review some early results on the SBM and its variants. First we introduce a consistency framework for community detection established by Bickel and Chen \cite{Bickel&Chen2009}. 
	They developed general theory for checking the consistency of a large class of community detection
	criteria under the SBM as the number of nodes $n$ grows and the number of communities $K$ remains fixed.
	
	For any label assignments $\V{e}$, let $O(\V{e})$ be a $K\times K$ matrix with entries $\{O_{kl}(\V{e}) \} $ defined by  
	$$O_{kl}(\V{e})=\sum_{1 \leq i,j \leq n} A_{ij}I\{e_i=k,e_j=l\} ,
	$$ 
	where $I$ is the indicator function.  And define
	$$D_{k}(\V{e})=\sum_{l=1}^K O_{kl}(\V{e}) ,\ \ L=\sum_{1 \leq i,j \leq n} A_{ij} . 
	$$ 
	For $k \neq l$, $O_{kl}$ is the number of edges between communities $k$ and $l$; $O_{kk}$ is twice the number of edges within community $k$; $D_{k}$ is the sum of node degrees in community $k$; and $L$ is the sum of all degrees in the whole network. 
	
	Define $n_k(\V{e})=\sum_{i=1}^n I\{e_i=k\}$ to be the number of nodes in community $k$, and $f(\V{e})=\left (n_1/n, n_2/n,...,n_K/n \right)$ to be the fractions of nodes in each community.

	A large class of community detection criteria can be written as the following general form up to a constant:
	\begin{align*}
	Q(\V{e})=F\left ( \frac{O(\V{e})}{\mu_n},\frac{L}{\mu_n},f(\V{e}) \right) ,
	\end{align*}
	where $\mu_n=\mathbb{E}(L)$. This class include many graph cut methods mentioned in the introduction such as the normalized cut \cite{Shi&Malik2000}, defined by
	\begin{align*}
	Q_{Ncut}(\V{e}) = -\sum_{k=1}^K \frac{D_k-O_{kk}}{D_k}.
	\end{align*}
	Newman-Girvan modularity \cite{NewmanPNAS}, defined by
	\begin{align*}
	Q_{NG}(\V{e}) = \sum_{k=1}^K \frac{O_{kk}}{L}- \left ( \frac{D_k}{L} \right )^2,
	\end{align*}
	also has this form. Moreover, Bickel and Chen also studied the profile likelihood of the SBM. If we treat community labels as fixed parameters, the log-likelihood of $A$ is 
	\begin{align*}
	\frac{1}{2} \sum_{1 \leq k,l\leq K} ( O_{kl} \log (P_{kl})+ (n_{kl}-O_{kl}) \log (1-P_{kl})),
	\end{align*}
	where $n_{kl}=n_{k}n_l$ if $k\neq l$ and $n_{kk}=n_k(n_k-1)$. In order to maximize the log-likelihood, we can first fix $\V{e}$ and maximize it over $P$. By doing so, we obtain the profile likelihood 
	\begin{align*}
	Q_{SBM} (\V{e}) = \sum_{1 \leq k,l\leq K} n_{kl} \tau \left ( \frac{O_{kl}}{n_{kl}} \right ), 
	\end{align*} 
	where $\tau(x)=x \log x +(1-x) \log (1-x)$. Bickel and Chen stated that $Q_{SBM}$ can also be written as the general form. 
	
	\begin{remark}
		The community labels $c_i$ are assumed to be fixed when the profile likelihood $Q_{SBM}$ is derived. But $c_i$ will be assumed to be random variables with $Multinomial(1,\pi)$ in Theorem \ref{Bickel}.  Similar phenomena are in fact very common in the study of community detection. It is worth emphasizing the difference between a model for theoretical analysis and a detection criterion for finding the partition in practice. A detection criterion can be derived from a model such as the SBM, or from a modified version of a model, or may not even be motivated by any model. In any case, it is worthwhile studying the consistency of this criterion. Bickel and Chen provided a general framework for this purpose. 
	\end{remark}
	
	Let $\hat{\V{c}}=\argmax_{\V{e}} Q(\V{e})$. A natural necessary condition for consistency of $\hat{\V{c}}$ is that the ``limit'' or ``population version'' of $Q(\V{e})$ should be maximized by the correct partition. We need more notations to specify this key condition.
	
	Define $\lambda_n=\mu_n/n$ to be the average expected degree and $\rho_n=\mu_n/(n(n-1))$ to be the expected graph density. Let $R$ be a $K \times K$ matrix with entries $\{ R_{ka} \}$ defined by 
	\begin{align*}
	R_{ka}=\frac{1}{n} \sum_{i=1}^n I(e_i=k,c_i=a) .
	\end{align*}
	$R_{ka}$ measures the fraction of nodes from community $a$ but classified into community $k$. Define $S_{ab}=P_{ab}/\rho_n$ for $1\leq a,b \leq K$. Note that $S_{ab}$ is independent of $n$.  
	
	Bickel and Chen stated the following condition:
	
	$F(RSR^T,1,R \V{1})$ is uniquely maximized over $\mathcal{R}=\{R: R\geq 0, R^T \V{1}=\pi \}$ by $R=\mathcal{D} (\pi )$, for all $(\pi,S)$ in an open set $\Theta$, where $\V{1}= (1,1,...,1)^T$ and $\mathcal{D} (\pi )$ is a diagonal matrix with $\pi$ as its diagonal elements.

	\begin{remark}
		Despite its seemingly complicated form, the key condition is very natural, following the same principle of $M$-estimators. In the authors' opinion, not only this condition can help researchers check the consistency of existing detection criteria, but it also provides guidance for designing new criteria. That is why we specify it in detail. 
		
	\end{remark}
	
	\begin{thm}[Theorem 1 in Bickel and Chen \cite{Bickel&Chen2009}]\label{Bickel}
		Suppose $F$, $S$ and $\pi$ satisfy the above condition and some mild regularity conditions. Suppose $\lambda_n/\log n \rightarrow \infty$. Then up to a permutation,
		$\mathbb{P}[\hat{\V{c}}=\V{c}] \rightarrow 1
		$.
	\end{thm}
	Bickel and Chen then applied Theorem \ref{Bickel} to study the consistency of the SBM profile likelihood and Newman-Girvan modularity. And other criteria such as the normalized cut can also be checked using the theorem. As one may expect, the SBM profile likelihood is consistent without additional parameter constraints, since the underlying model is the SBM. Even within-group densities are not required to be larger than the cross-group densities. By contrast, Newman-Girvan modularity requires such conditions to be consistent.

	\begin{remark}
		The result in Theorem \ref{Bickel} is called \textit{strong consistency} in statistics literature \cite{Zhaoetal2012,Amini.et.al.2013}, or \textit{exact recovery} in computer science literature \cite{Abbe16}. It requires no error in the estimated label vector with high probability, i.e. with probability approaching 1. During the proof, Bickel and Chen also obtained the result of weak consistency, that is, the fraction of misclassified nodes converging to 0, under a weaker condition $\lambda_n \rightarrow \infty$.   
	\end{remark}
	
	The SBM implies that nodes within a community have the same expected degree. But high-degree nodes, i.e. hubs do exist in many real-world networks \cite{Barabasi&Albert1999}. To address this issue, Karrer and Newman \cite{Karrer10} proposed the degree-corrected stochastic blockmodel (DCSBM), which allows more variation among node degrees within a community. Specifically, link probability in \eqref{SBM_prob} was replaced with $\mathbb{E}[A_{ij}|\V{c}]=\theta_i\theta_j P_{c_ic_j}$, where parameter $\theta_i$ controls the degree of node $i$. Zhao et al. \cite{Zhaoetal2012} generalized the framework of Bickel and Chen \cite{Bickel&Chen2009}  and obtain a general theorem for community detection consistency under the DCSBM. 
	
	The results \cite{Bickel&Chen2009,Zhaoetal2012} require that the number of communities remains as fixed. Choi et al. \cite{Choietal2011} established weak consistency 
	of the maximum likelihood estimator (MLE) under the SBM when the number of communities is allowed to grow with the network size. Specifically, weak consistency holds when the number of communities grows no faster than $n^{1/2}$, the average expected degree grows faster than $(\log n)^{3+\delta}$ for some $\delta>0$, and the minimum size of community is proportional to $n/K$. 
	
	\section*{\sffamily \Large PSEUDO-LIKELIHOOD, VARIATIONAL METHODS AND BELIEF PROPAGATION} 
	
	Many community detection criteria have good theoretical properties under the framework of SBM. However, the optimization of these criteria, including the maximum likelihood of SBM itself, is a great challenge in practice. As discrete optimization, finding global optimizers of these criteria requires the search over $K^n$ possible assignments, which is computationally intractable. 
	
	The expectation-maximization (EM) algorithm for fitting the likelihood of SBM faces the same difficulty. Unlike fitting the Gaussian mixture model, where the posterior probabilities of each cluster label can be calculated separately, the E-step for fitting SBM involves $K^n$ possible assignments \cite{Amini.et.al.2013}. This is due to the ``two-dimensional'' structure of networks as previously mentioned. We review two methods designed for overcoming this issue.
	
	Amini et al. \cite{Amini.et.al.2013} proposed a scalable pseudo-likelihood method for fitting the SBM and DCSBM, and proved consistency under the SBM with two communities. We adopt all the notation in the previous section and define a few more in order to introduce the method. Let $\V{e}$ be an initial labeling vector. Let $\V{b}_i$ be a vector of length $K$, with entries $\{b_{ik} \}$ defined by $b_{ik}=\sum_j A_{ij} I(e_i=k)$. $\V{b}_i$ are the block sums for column $i$. Amini et al. made the following observations: for each node $i$, conditional on $\V{c}$ with $c_i=l$:
	\begin{itemize}
		\item $\{b_{i1},b_{i2},...,b_{iK} \}$ are mutually independent;
		\item $b_{ik}$ is approximately Poisson with mean $\lambda_{lk}=n R_{k\cdot} P_{\cdot l}$. 
	\end{itemize}
	Amini et al. then proposed the pseudo-likelihood as follows (up to a constant),
	\begin{align*}
	\mathcal{L}_{PL}(\{\V{b}_i \})=\sum_{i=1}^n \log \left ( \sum_{l=1}^K \pi_l e^{-\lambda_l} \prod_{k=1}^K \lambda_{lk}^{b_{ik}} \right ), 
	\end{align*}
	where $\lambda_l=\sum_{k} \lambda_{lk}$. Amini et al. made several approximations to obtain the above pseudo-likelihood. First, the dependence among $\{\V{b}_i \}$ is ignored, which is reasonable since the dependence becomes very weak as $n$ grows but $K$ remains fixed. Second, Poisson approximation is used, which is also natural. Last, but most importantly, note that $\mathcal{L}_{PL}(\{\V{b}_i \})$ is not a likelihood of the original adjacency matrix $A$, but a likelihood of the block sums $\{\V{b}_i \}$, where $\V{b}_i$ depend on the initial labeling $\V{e}$. Therefore, the performance of this method can be sensitive to the accuracy of the initial labeling.  
	
	$\mathcal{L}_{PL}(\{\V{b}_i \})$ is the log-likelihood of a Poisson mixture model, and thereby the latent labels $\V{c}$ can be estimated by a standard EM algorithm. Note that now the posterior probabilities for $c_i$ can be calculated separately and thus very fast, since $\V{b}_i$ are independent. Once the EM algorithm converges, $\V{e}$ is updated to the
	most likely label for each node as indicated by the EM and the procedure repeats a fixed number of iterations. Amini et al. proposed a pseudo-likelihood conditional on node degrees (CPL) and developed a similar algorithm for fitting the DCSBM. 
	
	Amini et al. proved the weak consistency of the estimator from one-step EM of CPL for $K=2$ under the SBM. We omit the details of the estimator since it would require a lot more complicated notation otherwise. True community labels $\V{c}$ are treated as fixed parameters. For simplicity, we only present the result for balanced communities, i.e., each community contains $m=n/2$ nodes.  Assume the link probability matrix $P$ has the form 
	\begin{align*}
	P=\frac{1}{m}\left(
	\begin{array}{cc}
	a & b  \\
	b & a  \\
	\end{array}
	\right). 
	\end{align*}
	Let $\hat{a}$ and $\hat{b}$  be some initial estimates of $a$ and $b$. And assume that the initial labeling is balanced and it matches exactly $\gamma m$ labels in community 1. 
	\begin{thm}[Theorem 2 in Amini et al.\cite{Amini.et.al.2013}]\label{AAA}
		The one-step EM estimator of CPL is weakly consistent under some mild regularity conditions and the following main assumptions:
		\begin{itemize}
			\item [(C1)] $\gamma \neq 1/2$;
			\item [(C2)] $(\hat{a}-\hat{b})(a-b)>0$;
			\item [(C3)] $(a-b)^2/(a+b) \rightarrow \infty$.
		\end{itemize}
	\end{thm}
	All these assumptions are intuitive and very mild. Condition (C1) only requires the initial labeling better than random guessing. Condition (C2) means that the estimates $(\hat{a},\hat{b})$ should have the same ordering as true parameters $(a,b)$. And it is easy to check that $\lambda_n \rightarrow \infty$ implies (C3). On the other hand, it is worth noting that Theorem \ref{AAA} only guarantees consistency for the case of two communities. The proof is already highly technical and relies on advanced probability tools. It may be quite challenging to prove or even formulate the theorem for the general case. It is worth mentioning that Zhang and Zhou \cite{zhang2016minimax} proved that $(a-b)^2/a \rightarrow \infty$ is a necessary and sufficient condition for weak consistency when $a>b$ by providing a minimax theory for community detection. This result further justifies (C3). Gao et al. \cite{gao2015achieving} proposed a refinement scheme by adding a majority vote step to spectral clustering (to be introduced in the next section) which can achieve the minimax rate. The results were generalized into the DCSBM by Gao et al. \cite{gao2016community} 
	
	Daudin et al. \cite{Daudinetal2008} introduced a variational approach to overcome the computational challenge of the EM algorithm for fitting the SBM (see Tzikas et al. \cite{Tzikas2008} for a tutorial of variational approaches in general). We again adopt the notation in the previous section when introducing this approach. Further, Let $Z=[z_{ik}]$ be an $n \times K$ matrix, where $z_{ik}=1$ if $c_i=k$. Here $Z_i=(z_{i1},z_{i2},...,z_{iK})$ follows $Multinomial(1,\pi)$. Let $R_A(Z)$ be a function of $Z$, which depends on the adjacency matrix $A$. Define 
	\begin{align}\label{variational}
	\mathcal{T}(R_A; \pi,P)=\log \mathcal{L}(A;\pi,P)-KL[R_A(\cdot),\mathbb{P}(\cdot|A;\pi,P)],
	\end{align}
	where $KL$ denotes the Kullback–Leibler divergence, $\mathcal{L}(A;\pi,P)$ is the marginal log-likelihood of $A$ and $\mathbb{P}(Z|A;\pi,P)]$ is the posterior probability for community labels. Note that if we put no constraint on $R_A$, then $\max_{\pi,P,R_A} \mathcal{T}(R_A(Z); \pi,P)=\max_{\pi, P} \log \mathcal{L}(A;\pi,P)$, since taking $R_A(\cdot)=\mathbb{P}(\cdot|A;\pi,P)$ makes the second term of \eqref{variational} disappear. According to this observation, the EM algorithm can be viewed as two alternating maximization steps: in order to maximize $\mathcal{T}(R_A; \pi,P)$, the algorithm alternately solves for $(\pi,P)$ given $R_A$, which is the M step, and solves for $R_A$ given $(\pi,P)$, which is the E step (see Hastie et al. \cite{hastie01} and Tzikas et al. \cite{Tzikas2008} for details). 
	
	As mentioned earlier, it is intractable to compute $\mathbb{P}(Z|A;\pi,P)]$. The key idea of the variational approach in Daudin et al. \cite{Daudinetal2008} is to replace $\mathbb{P}(Z|A;\pi,P)]$ by a tractable $R_A(Z)$. They constraint $R_A(Z)$ to have the form $R_A(Z)=\prod_i h(Z_i;\tau_i) $, where $\tau_i=(\tau_{i1},...,\tau_{iK})$ and $h(\cdot; \tau)$ denotes the multinomial distribution with parameter $\tau$. Note that $R_A(Z)$ is a product and is given a parametric form with unknown parameters $\tau_i$. Now, parameters $\tau_i$, $\pi$ and $P$ can be iteratively updated, following the same procedure in the last paragraph. 
	
	Celisse et al. \cite{Celisseetal2011} established the consistency of the variational estimators for parameters $(\pi, P)$ in the SBM, in which the expected graph density $\rho_n$ is fixed.  Bickel et al. \cite{Bickel&Choi&etal2012} established the consistency and asymptotic normality of the variational estimators, in which $\rho_n$ can go to 0.
	
	Belief propagation, as an algorithm for inference on graphical models, was also applied in community detection by researchers \cite{hastings2006community,mossel2014belief,mossel2015reconstruction,mossel2015density,mossel2016local}. We refer the reader to Yedidia et al. \cite{yedidia2003understanding} for a tutorial introduction to the classical belief propagation method for graphical models such as Bayesian networks and Markov random fields. We now focus on a specific belief propagation algorithm for community detection proposed by Mossel and Xu \cite{mossel2015density}. 
	
	As mentioned earlier, label assignments, i.e., the E step is computationally infeasible for the SBM. Belief propagation is an alternative approach to assigning community labels approximately but efficiently given the parameters in the SBM. Mossel and Xu \cite{mossel2015density} proposed a belief propagation algorithm for the SBM with two communities and known $(\pi,P)$. When the parameters are unknown, the algorithm can be used as the E step in the EM algorithm. Assume that the link probability matrix $P$ has the form 
	\begin{align*}
	P=\frac{1}{n}\left(
	\begin{array}{cc}
	a & b  \\
	b & c  \\
	\end{array}
	\right). 
	\end{align*}
	Let $\partial i$ be the set of neighbors of $i$ and $F(x)=\frac{1}{2} \log \left ( \frac{e^{2x}\pi_1 a +\pi_2 b}{e^{2x}\pi_1 b +\pi_2 c} \right )$. Let $d_+=\pi_1 a +\pi_2 b$ and $d_-= \pi_1 b +\pi_2 c $. At $t$-th iteration, define
	\begin{align*}
	R^t_{i \rightarrow j}=\frac{-d_+ + d_-}{2}+\sum_{l \in \partial i \backslash \{ j\}} F(R^{t-1}_{l \rightarrow i}),
	\end{align*}
	which can be intuitively understood as a message from node $i$ to node $j$ about which community node $j$ should belong to. And the belief of node $u$ at $t$-th iteration $R^t_u$ is defined as
	\begin{align*}
	R^t_u=\frac{-d_+ + d_-}{2}+\sum_{l \in \partial u} F(R^{t-1}_{l \rightarrow u}),
	\end{align*} 
	which is an approximation of $\frac{1}{2} \log \frac{\mathbb{P} (A|c_u=1)}{\mathbb{P} (A|c_u=2)}$. And thus label assignments can be easily determined by $R^t_u$. 
	
	\textbf{Algorithm 1:} (Belief propagation for community detection \cite{mossel2015density})
	\begin{itemize}
		\item [(1)] Set $R^0_{i \rightarrow j}=0$. 
		\item [(2)] Compute $R^t_{i \rightarrow j}$ for $T-1$ iterations.
		\item [(3)] Compute $R^T_{i}$ for all $i=1,...,n$. 
		\item [(4)] Return $\hat{c}_i= 2-I(R^T_i \geq -\psi)$ for all $i=1,...,n$, where $\psi=\frac{1}{2} \log \frac{\pi_1}{\pi_2}$. 
	\end{itemize}
	
	We now give a brief explanation of why Algorithm 1 works especially for sparse networks. This algorithm gives an exact solution for tree models defined as follows. 
	\begin{definition}[Definition 3.1 in Mossel and Xu \cite{mossel2015density}] 
		For a node $u$, denote by $(T_u,\V{c})$ the following Poisson two-type branching process tree rooted at $u$, where $\V{c}$ are the labels of the nodes of $T_u$. Let $c_u=1$ with probability $\pi_1$ and $c_u=2$ with probability $\pi_2$. Recursively for each node $i$ in $T_u$, given $c_i=1$, $i$ will have $Pois(\pi_1 a)$ children $j$ with $c_j=1$ and $Pois(\pi_2 b)$ children $j$ with $c_j=2$; given $c_i=2$, $i$ will have $Pois(\pi_1 b)$ children $j$ with $c_j=1$ and $Pois(\pi_2 c)$ children $j$ with $c_j=2$.
	\end{definition}
	The belief $R^t_u$ is an exact solution for $\frac{1}{2} \log \frac{\mathbb{P} (A|c_u=1)}{\mathbb{P} (A|c_u=2)}$ if $A$ is such a tree of depth $t$ rooted at node $u$ \cite{mossel2015density}. The remaining question is why $A$ generated by the SBM can be approximated by a tree defined above. First, note that when $A$ is sparse, its structure can be similar to a tree. Second, under the SBM, node $i$ is connected with $Bin(n-1,\pi_1 a/n)$ nodes $j$ with $c_j=1$ given $c_i=1$. Thus according to Poisson approximation to Binomial, node $i$ is connected with approximate $Pois(\pi_1 a)$ nodes $j$ with $c_j=1$, which is consistent with the above definition. Similar results hold for other cases. Mossel and Xu \cite{mossel2015density} obtained an asymptotic formula for the fraction of mis-classified nodes on average by Algorithm 1 and proved that it achieves the minimum mis-classification rate. 
	
	\section*{\sffamily \Large SPECTRAL CLUSTERING APPROACHES} 
	Both pseudo-likelihood and variational approaches require initial values and their performance can be sensitive to the accuracy of initial values. In this section, we review another class of computationally feasible approaches -- spectral clustering, using eigenvectors of adjacency matrices or graph Laplacian matrices (defined later in this section), which do not require initial values. Spectral clustering has a long history. The algorithm and its variations have been applied into different fields. We refer the reader to von Luxburg \cite{vonLuxburg2007} for a tutorial. 
	
	Theoretical properties for variants of spectral clustering for community detection has been studied by a number of researchers. Rohe et al. \cite{Rohe2011} studied the asymptotic behavior of spectral clustering under the SBM. Chaudhuri et al. \cite{Chaudhuri&Chung&Tsiatas2012} introduced a degree-corrected graph Laplacian for the extended planted partition model. Qin and Rohe \cite{Qin2013} applied a regularized graph Laplacian matrix into the traditional spectral clustering algorithm and gave the bound for mis-classification rate under the DCSBM. Fishkind et al. \cite{Fishkind13} established the consistency of a modified spectral clustering procedure, which only requires the knowledge of an upper bound on the number of communities. Sarkar and Bickel \cite{Sarkar&Bickel2015} compared the asymptotic behavior of normalized and unnormalized spectral clustering for the SBM. Jin \cite{Jin2015} proposed spectral clustering on ratios-of-eigenvectors (SCORE) for the DCSBM. Lei and Rinaldo \cite{Lei&Rinaldo2015} established the consistency of spectral clustering under the SBM where the order of the maximum expected degree is $\log(n)$. Most spectral clustering methods are based on the adjacency matrix or the graph Laplacian and their variants. Besides that, other matrices are used for spectral clustering. For instance, Krzakala et al. \cite{Krzakala13} used the non-backtracking matrix for community detection in sparse networks. Le and Levina \cite{Le2015} considered the estimation of the number of communities that uses spectral properties of the Bethe Hessian matrix and the non-backtracking matrix.
	
	Next, we briefly review the methods in Rohe et al. \cite{Rohe2011} and Jin \cite{Jin2015} to provide some insight into why spectral clustering works for community detection. 
	
	Rohe et al. \cite{Rohe2011} studied spectral clustering with the normalized graph Laplacian. Let $D$ be a $n\times n$  diagonal matrix with $D_{ii}=\sum_j A_{ij}$. The normalized graph Laplacian is defined as $L=I-D^{-1/2} A D^{-1/2}$. Rohe et al. in fact considered $L=D^{-1/2} A D^{-1/2}$, but it makes no difference in eigen-analysis. 
	
	\textbf{Algorithm 2:} (Spectral clustering based on graph Laplacian \cite{Rohe2011,vonLuxburg2007})
	\begin{itemize}
		\item [(1)] Find the eigenvectors $X_1,...,X_K$ corresponding to the $K$ eigenvalues of $L$ with largest absolute values. Define $X=[X_1,...,X_K]$ by putting the eigenvectors into the columns. 
		\item [(2)] Treating each of the $n$ rows in $X$ as a point in $\mathcal{R}^K$, denoted by $X'_1,...,X'_n$, run $k$-means with $K$ clusters. This creates a disjoint partition of $V$ into $K$ communities. 
	\end{itemize}
	$k$-means is a classical clustering method in multivariate analysis (see Hastie et al. \cite{hastie01} or other textbooks on machine learning for details), which optimizes the following criterion 
	$$\min_{m_1,...,m_K,V_1,...,V_K } \sum_{k=1}^K \sum_{i \in V_k} \| X'_i-m_k \|^2, $$
	where $\{V_1,...,V_K\}$ forms a disjoint partition of $V$. 
	
	Spectral clustering transforms a community detection problem to a clustering problem by eigen-decomposition. The rationale behind this approach can be explained by the idea of ``population version'' mentioned in the second section. If we adopt the notation $Z$ introduced in the previous section, and treat it as fixed, then the population version $A$ is $\mathscr{A}=Z P Z^T$. In fact, $\mathbb{E}[A_{ij}]=\mathscr{A}_{ij}$ except for the diagonal elements, whose effect is yet very minor. Perform the spectral clustering on this population version $\mathscr{A}$. That is, let $\mathscr{L}$ be the graph Laplacian of $\mathscr{A}$. Further, let $\mathscr{X}=[\mathscr{X}_1,...,\mathscr{X}_K]$, of which columns are the eigenvectors corresponding to the nonzero eigenvalue of $\mathscr{L}$. It is easy to prove that there are $K$ unique rows in $\mathscr{X}$, which implies a perfect community partition in the sense of population version \cite{Rohe2011}. Furthermore, one can expect that the rows of the ``noisy version'' $X$ concentrate around the $K$ \textit{centroids} and hence can be clustered by $k$-means. Rohe et al. gave a bound for the number of mis-classified nodes under the SBM. In particular, Rohe et al. studied the planted partition model with equal sized communities as an example. The planted partition model, denoted by $\mathcal{G}(n,p,q)$, is a special case of the SBM, where the diagonal elements of $P$ are a constant $p$ and the off-diagonal elements are another constant $q$. Rohe et al. showed that under the planted partition model with equal sized communities, the mis-classification rate is $o(n^{-1/4})$ almost surely, when $k=O(n^{1/4}/\log n)$ and $p,q$ remain as fixed. 
	
	Jin \cite{Jin2015} proposed the spectral clustering on ratios-of-eigenvectors (SCORE) which is designed for DCSBM. First find the eigenvectors corresponding to the $K$ eigenvalues of $A$ with largest absolute values: $\hat{\eta}_1=[\hat{\eta}_{11},....,\hat{\eta}_{1n}],\hat{\eta}_2=[\hat{\eta}_{21},....,\hat{\eta}_{2n}],...,\hat{\eta}_K=[\hat{\eta}_{K1},....,\hat{\eta}_{Kn}]$. And Let $\hat{R}^{*}$ be an $n\times (K-1)$ matrix  with entries defined by 
	\begin{align*}
	\hat{R}^{*}_{ik}=\begin{cases}
	\hat{R}_{ik} & $if $ |\hat{R}_{ik}|\leq \log n , \\
	\log n & $if $ \hat{R}_{ik}> \log n , \\
	-\log n & $if $ \hat{R}_{ik}< -\log n , \\
	\end{cases} 				
	\end{align*}
	where $\hat{R}_{ik}=\hat{\eta}_{Ki}/\hat{\eta}_{ki}$. Finally, run $k$-means on rows of $\hat{R}^{*}$ to obtain community labels. Jin proved that the SCORE is weakly consistent under the DCSBM when $K$ remains fixed.
	\begin{remark}
		When proving consistency, both Rohe et al. \cite{Rohe2011} and Jin \cite{Jin2015} in fact considered the global optimizer of $k$-means. However, the global optimization of $k$-means is NP-hard. Lei and Rinaldo \cite{Lei&Rinaldo2015}  considered an approximate $k$-means algorithm solvable in polynomial time \cite{Kumar2004} and proved the consistency of spectral clustering with this algorithm under the SBM. 
	\end{remark}
	
	Spectral clustering based on the standard graph Laplacian is known to perform poorly on sparse graphs \cite{Amini.et.al.2013,Cai15,AAA16,Le16}, i.e., graphs with link probabilities of order $1/n$.  The problem lies with low-degree nodes that can cause irregular behavior of the graph Laplacian. A regularized graph Laplacian was proposed by Amini et al. \cite{Amini.et.al.2013} and studied in several articles \cite{joseph2013impact,le2017concentration}. Specifically, we replace the adjacency matrix $A$ with $A_\tau=A+(\tau/n)\V{1} \V{1}^T$ and construct the graph Laplacian using $A_\tau$, where $\tau$ is a quantity with the same order of the average expected degree. Le et al. \cite{le2017concentration} proved that for the planted partition model $\mathcal{G}(n,a/n,b/n)$, spectral clustering with the regularized graph Laplacian correctly estimates the communities up to at most $\epsilon n$ mis-classified nodes, if $(a-b)^2> C_\epsilon (a+b)$ where $C_\epsilon$ is a constant depending on $\epsilon$.

	\section*{\sffamily \Large SEMIDEFINITE PROGRAMMING FOR THE SBM} 
	
	As shown in the previous section, spectral clustering algorithms are usually neat and easy to implement. And their theoretical performance was also justified in literature. Therefore, spectral clustering approaches are very promising  from
	both  a  theoretical  and  computational  point  of  view, as pointed out by Bickel et al. \cite{Bickel&Choi&etal2012}. On the other hand, some limitation of spectral clustering was pointed out by very recent literature \cite{Cai15,AAA16,Le16}. These authors argued that spectral clustering works well for dense networks but may fail for sparse networks. Besides, some spectral clustering algorithms can be viewed as non-convex relaxations of certain graph cut criteria \cite{Shi&Malik2000,Newman2006}, and usually rely on $k$-means as the final step to get discrete labels. As mentioned in the previous section, the global optimization of $k$-means is however NP-hard and the commonly used algorithm can only guarantee local solutions. Therefore, some researchers are interested in convex relaxations, more specifically,  semidefinite programming (SDP) relaxations for these criteria. 
	
	Recently, SDP approaches to fitting the SBM or its variants have been proposed in the literature \cite{Chen_NIPS,Chen2016,Cai15,AAA16,Sarkar16,Abbe16}. Gu{\'e}don and Vershynin \cite{guedon2016community} developed a general method to prove consistency of SDP by Grothendieck's inequality and proved that various SDP methods can recover the community structure up to an arbitrarily small fraction
	of mis-classified nodes in sparse graphs.   
	
	Here we review some neat results in a very recent published article \cite{Abbe16}. Abbe et al. \cite{Abbe16} was interested in sharp threshold for exact recovery of communities under the SBM. In particular, they considered the simplest case of the SBM -- the planted partition model (defined in the previous section) with two equal sized communities, denoted by $\mathcal{G}(n,p,q)$. Letting $\alpha=p n/\log n$ and $\beta=q n/\log n$, and assuming $\alpha,\beta$ are constant and $\alpha>\beta$, Abbe et al. proved the following result:
	\begin{thm}[Theorem 1 and 2 in Abbe et al. \cite{Abbe16}]
		If $(\alpha+\beta)/2-\sqrt{\alpha \beta}>1$, then the MLE of $\mathcal{G}(n,p,q)$ exactly recovers the communities (up to a permutation), with high probability. 
		
		Conversely, if $(\alpha+\beta)/2-\sqrt{\alpha \beta}<1$, then for sufficiently large $n$, the MLE fails in recovering the communities
		with probability bounded away from zero.
	\end{thm} 
	
	Therefore, $(\alpha+\beta)/2-\sqrt{\alpha \beta}$ is a sharp threshold for exact recovery by the MLE. This result is stronger than the one in Bickel and Chen \cite{Bickel&Chen2009} in the sense that it allows the average expected degree $\lambda_n$ to have order $\log n$, while Bickel and Chen \cite{Bickel&Chen2009} requires $\lambda_n/\log n \rightarrow \infty$. On the other hand, Bickel and Chen \cite{Bickel&Chen2009} allows an arbitrary number of communities $K$. 
	
	As mentioned earlier, solving the MLE of the SBM is computationally infeasible. Abbe et al. then proposed a SDP approach which can exactly recover the communities when $\lambda_n=\Theta(\log n)$. Define $g=(g_1,....,g_n)^T$, where $g_i=+1$ if node $i$ belongs to the first community and $g_i=-1$ if node $i$ belongs to the second community. Further, define $B$ as an $n\times n$ matrix with zero diagonal whose off-diagonal elements $B_{ij}=2A_{ij}-1$. The following criterion aims  to   find two
	communities such that the number of within-community edges
	minus the cross-community edges is maximized:
	\begin{align}\label{discrete}
	\max_{g} & g^T B g  \nonumber \\
	\mbox{s.t. } & g_i=\pm 1. 
	\end{align}
	Abbe et al. proposed the following SDP relaxation for \eqref{discrete}, which can be solved in  polynomial time.
	\begin{align}\label{SDP}
	\max_{X \in \mathcal{R}^{n \times n}} & Tr(BX)  \nonumber \\
	\mbox{s.t. } & X_{ii}=1 \\
	&  X \succeq 0,
	\end{align}
	where $X \succeq 0$ means that $X$ is positive-semidefinite. Abbe et al. proved the following result:
	\begin{thm}[Theorem 3 in Abbe et al. \cite{Abbe16}]
		If $(\alpha-\beta)^2>8(\alpha+\beta)+\frac{8}{3}(\alpha-\beta) $, the following holds with high probability: \eqref{SDP} has a unique solution which is given by the outer-product of $g \in \{\pm 1\}^n$ whose  entries
		corresponding to the first community are 1 and to the second community are -1.
	\end{thm}
	
	A related but different concept is weak discovery, also called detection. Weak discovery only requires the algorithm to find a
	partition which is positively correlated with the true communities with high probability. Decelle et al. \cite{Decelle2011} made a remarkable conjecture on the threshold of weak discovery for the planted partition model based on deep ideas from statistical physics. Specifically, let $a=p n$ and $b=q n$. Then Decelle et al. \cite{Decelle2011} conjectured that it is possible to develop a polynomial-time algorithm to achieve weak discovery if $(a-b)^2>2(a+b)$ and is impossible if $(a-b)^2<2(a+b)$. The conjecture for the case of two symmetric communities was proved independently by Massouli{\'e} \cite{Massoulie2014} and Mossel et al. \cite{mossel2013proof} Physicists \cite{nadakuditi2013spectra,zhang2014} also considered the threshold of weak discovery for networks with arbitrary degrees. 
	
	\section*{\sffamily \Large OTHER TOPICS ON COMMUNITY DETECTION} 
	
	In this section, we briefly review other topics on community detection related to consistent detection methods under the SBM. These research fields are nascent compared to the study of the SBM. Therefore, some of the methods introduced in this section may have been developed intuitively without theoretical justification. 
	
	\subsection*{\sffamily \large ROBUST COMMUNITY DETECTION}
	
	The SBM makes strong assumptions on networks, that is, every node is assumed to belong to a homogeneous block. However, many real-world networks contain ``outliers'', that is, nodes that do not
	fit in with any of the communities. Therefore, robust community detection methods are desirable in real applications. The term of robust community detection is not well defined, and there is no agreement on its scope. We focus on detection methods robust to outliers as described above. Zhao et al. \cite{Zhao.et.al.2011} proposed a sequential approach called community extraction, which extracts one community at a time,  allowing for arbitrary structure in the remainder of the network. 
	At each step, the extraction criterion looks for a cohesive group with more links within itself than to the rest of the network, but ignores links within its complement. Cai and Li \cite{Cai15} proposed the generalized stochastic blockmodel that allows for outliers to be connected with the other nodes  in  the  network  in  an  arbitrary  way, and fitted the model by SDP. The notion of outliers in Cai and Li \cite{Cai15} is different from the one in Zhao et al. \cite{Zhao.et.al.2011}: the link pattern between a community and outliers is also arbitrary.  Another class of robust methods is local community detection \cite{Flake02,Clauset2005,Wu_local15,Laarhoven_local16,Qi_local2014}. Instead of partitioning the entire network into communities, local community  detection methods  seek  a single community  of  nodes
	concentrated around a few given seed nodes, based on certain criteria measuring cohesiveness of a group such as conductance \cite{Laarhoven_local16}. This technique is particularly useful when the network is not completely known and only local information is available. To the best of our knowledge, no theoretical framework has been established for local community detection, which is a possible direction for future research.
	
	\subsection*{\sffamily \large COMMUNITY DETECTION WITH NODAL COVARIATES}
	Traditional community detection approaches only use the adjacency matrix, i.e. the network itself as the input. However, additional information on the nodes are usually available in addition to network topology. Thus a natural question is how or whether we can improve community detection by using node features, when presumably these features are correlated to community structure. Recently there have been a number of works on community detection with nodal covariates. Binkiewicz et al. \cite{Rohe2014} modified spectral clustering with the help of nodal covariates and justified the proposed method under the so-called   node-contextualized stochastic blockmodel. Zhang et al. \cite{Zhang2016} proposed a joint community detection criterion that
	uses both the adjacency matrix and nodal covariates by weighing edges according to nodal similarities. Yan and Sarkar \cite{Sarkar16} combined a similarity matrix based on nodal covariates with the adjacency matrix in a SDP problem. Furthermore, likelihoods of link probabilities incorporating auxiliary nodal information were proposed in literature \cite{Xu2012,Yang13,Newman2016,Handcock2007,Hoff2009}. In the author's opinion, a particular challenge in community detection with nodal covariates is how to assess whether or not covariates are correlated with the community structure induced by the adjacency matrix. Sometimes, covariates and the network showed different community structures. Even when they are correlated, it is not clear whether combining them is necessarily better than using only one source. 
	Yan and Sarkar \cite{Sarkar16} provided an answer along this line of thinking. But clearly more research can be conducted for this question.
	
	\subsection*{\sffamily \large DETERMINING THE NUMBER OF COMMUNITIES}
	
	Most of the methods we discussed so far require prior knowledge of the number of communities $K$. Even though, many asymptotic results allow the number of communities $K$ to grow with $n$, it is challenging to estimate this number in practice. Some methods have been proposed in recent years. Zhao et al. \cite{Zhao.et.al.2011} sequentially extracted communities until the rest of the network performed like an Erd{\"o}s-R\'enyi random graph based on a hypothesis test. Bickel and Sarkar \cite{Bickel:Sarkar:2015} designed a hypothesis test for the SBM based on the principal eigenvalue of a standardized adjacency matrix. Lei \cite{Lei:2016} proposed a goodness-of-fit test for the SBM based on the largest singular value of a residual matrix obtained by subtracting the estimated block mean effect from the adjacency matrix. The two approaches above rely on deep results in random matrix theory. Furthermore, BIC based approaches have been proposed in literature \cite{saldana2016many,Bickel:Wang2016,Hu2016}.
	
	\subsection*{\sffamily \large FUTURE RESEARCH} 
	We close our discussion with suggestions for future research. Firstly, current theoretical studies on community detection mainly focus on the SBM and its variants. According to the author's personal experiences, the SBM is not robust to ill-behaved nodes despite its theoretical convenience. Building theoretical frameworks for other models such as latent space models could be of interest to researchers. In particular, it seems to be natural to incorporate nodal covariates into latent space models. Secondly, community detection for weighted networks is an open problem. Lots of graph cut criteria can be applied to weighted networks. But model-based approaches with theoretical justification are desirable. Thirdly, developing community detection methods robust to outliers deserves further research efforts.


\begin{thebibliography}{10}
	
	\bibitem{Abbe16}
	E.~Abbe, A.~S. Bandeira, and G.~Hall.
	\newblock Exact recovery in the stochastic block model.
	\newblock {\em IEEE Transactions on Information Theory}, 62(1):471--487, 2016.
	
	\bibitem{Airoldi2008}
	E.~M. Airoldi, D.~M. Blei, S.~E. Fienberg, and E.~P. Xing.
	\newblock Mixed membership stochastic blockmodels.
	\newblock {\em J. Machine Learning Research}, 9:1981--2014, 2008.
	
	\bibitem{Albert&Barabasi2002}
	R.~Albert and A.-L. Barab\'{a}si.
	\newblock Statistical mechanics of complex networks.
	\newblock {\em Rev. Mod. Phys.}, 74:47--97, 2002.
	
	\bibitem{Amini.et.al.2013}
	A.A. Amini, A.~Chen, P.J. Bickel, and E.~Levina.
	\newblock Fitting community models to large sparse networks.
	\newblock {\em Annals of Statistics}, 41(4):2097--2122, 2013.
	
	\bibitem{AAA16}
	A.A. Amini and E.~Levina.
	\newblock On semidefinite relaxations for the block model.
	\newblock 2016.
	\newblock arXiv:1406.5647v3.
	
	\bibitem{Barabasi&Albert1999}
	A.-L. Barab\'{a}si and R.~Albert.
	\newblock Emergence of scaling in random networks.
	\newblock {\em Science}, 286:509--512, 1999.
	
	\bibitem{Bickel&Chen2009}
	P.~J. Bickel and A.~Chen.
	\newblock A nonparametric view of network models and {N}ewman-{G}irvan and
	other modularities.
	\newblock {\em Proc. Natl. Acad. Sci. USA}, 106:21068--21073, 2009.
	
	\bibitem{Bickel&Choi&etal2012}
	P.~J. Bickel, D.~Choi, X.~Chang, and H.~Zhang.
	\newblock Asymptotic normality of maximum likelihood and its variational
	approximation for stochastic blockmodels.
	\newblock {\em Annals of Statistics}, 41:1922--1943, 2013.
	
	\bibitem{Bickel:Sarkar:2015}
	P.~J. Bickel and P.~Sarkar.
	\newblock Hypothesis testing for automated community detection in networks.
	\newblock {\em Journal of the Royal Statistical Society: Series B},
	78:253--273, 2015.
	
	\bibitem{Rohe2014}
	N.~Binkiewicz, J.~T. Vogelstein, and K.~Rohe.
	\newblock Covariate-assisted spectral clustering.
	\newblock 2014.
	\newblock arXiv:1411.2158.
	
	\bibitem{Cai15}
	T.~Tony Cai and X.~Li.
	\newblock Robust and computationally feasible community detection in the
	presence of arbitrary outlier nodes.
	\newblock {\em Annals of Statistics}, 43(3):1027--1059, 2015.
	
	\bibitem{Celisseetal2011}
	A.~Celisse, J.-J. Daudin, and L.~Pierre.
	\newblock Consistency of maximum-likelihood and variational estimators in the
	stochastic block model.
	\newblock {\em Electronic Journal of Statistics}, 6:1847--1899, 2012.
	
	\bibitem{Chaudhuri&Chung&Tsiatas2012}
	K.~Chaudhuri, F.~Chung, and A.~Tsiatas.
	\newblock Spectral clustering of graphs with general degrees in the extended
	planted partition model.
	\newblock {\em JMLR Workshop and Conference Proceedings}, 23:35.1 -- 35.23,
	2012.
	
	\bibitem{Chen_NIPS}
	Y.~Chen, S.~Sanghavi, and H.~Xu.
	\newblock Clustering sparse graphs.
	\newblock In F.~Pereira, C.~J.~C. Burges, L.~Bottou, and K.~Q. Weinberger,
	editors, {\em Advances in Neural Information Processing Systems 25}, pages
	2204--2212. Curran Associates, Inc., 2012.
	
	\bibitem{Chen2016}
	Y.~Chen and J.~Xu.
	\newblock Statistical-computational tradeoffs in planted problems and submatrix
	localization with a growing number of clusters and submatrices.
	\newblock {\em Journal of Machine Learning Research}, 17:1--57, 2016.
	
	\bibitem{Choietal2011}
	D.~S. Choi, P.~J. Wolfe, and E.~M. Airoldi.
	\newblock Stochastic blockmodels with growing number of classes.
	\newblock {\em Biometrika}, 99:273--284, 2012.
	
	\bibitem{Clauset2005}
	A.~Clauset.
	\newblock Finding local community structure in networks.
	\newblock {\em Physical Review E}, 72(2):026132, 2005.
	
	\bibitem{Daudinetal2008}
	J.-J. Daudin, F.~Picard, and S.~Robin.
	\newblock A mixture model for random graphs.
	\newblock {\em Statist. Comput.}, 18:173--183, 2008.
	
	\bibitem{Decelle2011}
	A.~Decelle, F.~Krzakala, C.~Moore, and L.~Zdeborov{\'a}.
	\newblock Asymptotic analysis of the stochastic block model for modular
	networks and its algorithmic applications.
	\newblock {\em Physical Review E}, 84(6):066106, 2011.
	
	\bibitem{Erdos59}
	P.~Erd{\"o}s and A.~R{\'e}nyi.
	\newblock {On Random Graphs. I}.
	\newblock {\em Publicationes Mathematicae}, 6:290–--297, 1959.
	
	\bibitem{Fishkind13}
	D.~E. Fishkind, D.~L. Sussman, M.~Tang, J.~T. Vogelstein, and
	C.~E. Priebe.
	\newblock Consistent adjacency-spectral partitioning for the stochastic block
	model when the model parameters are unknown.
	\newblock {\em SIAM J. Matrix Anal. Appl.}, 34(1):23–39, 2013.
	
	\bibitem{Flake02}
	G.~M. Flake, S.~Lawrence, C.~L. Giles, and F.~M. Coetzee.
	\newblock Self-organization and identification of web communities.
	\newblock {\em IEEE Computer}, 35:66--71, 2002.
	
	\bibitem{Fortunato2010}
	S.~Fortunato.
	\newblock Community detection in graphs.
	\newblock {\em Physics Reports}, 486(3-5):75 -- 174, 2010.
	
	\bibitem{Fortunato16}
	S.~Fortunato and D.~Hric.
	\newblock Community detection in networks: A user guide.
	\newblock 2016.
	
	\bibitem{Fraley02}
	C.~Fraley and A.~E. Raftery.
	\newblock Model-based clustering, discriminant analysis, and density
	estimation.
	\newblock {\em Journal of the American Statistical Association},
	97(458):611--631, 2002.
	
	\bibitem{Frank&Strauss1986}
	O.~Frank and D.~Strauss.
	\newblock {M}arkov graphs.
	\newblock {\em Journal of the American Statistical Association}, 81:832--842,
	1986.
	
	\bibitem{gao2015achieving}
    C.~Gao, Z.~Ma, A.~Y Zhang, and H.~H Zhou.
	\newblock Achieving optimal misclassification proportion in stochastic block
	model.
	\newblock {\em arXiv:1505.03772}, 2015.
	
	\bibitem{gao2016community}
	C.~Gao, Z.~Ma, A.~Y Zhang, and H.~H Zhou.
	\newblock Community detection in degree-corrected block models.
	\newblock {\em arXiv:1607.06993}, 2016.
	
	\bibitem{Getoor2005}
	L.~Getoor and C.~P. Diehl.
	\newblock Link mining: A survey.
	\newblock {\em ACM SIGKDD Explorations Newsletter}, 7(2):3--12, 2005.
	
	\bibitem{Goldenberg2010}
	A.~Goldenberg, A.~X. Zheng, S.~E. Fienberg, and E.~M. Airoldi.
	\newblock A survey of statistical network models.
	\newblock {\em Foundations and Trends in Machine Learning}, 2:129--233, 2010.
	
	\bibitem{guedon2016community}
	O.~Gu{\'e}don and R.~Vershynin.
	\newblock Community detection in sparse networks via grothendieck’s
	inequality.
	\newblock {\em Probability Theory and Related Fields}, 165(3-4):1025--1049,
	2016.
	
	\bibitem{Handcock2007}
	M.~D. Handcock, A.~E. Raftery, and J.~M. Tantrum.
	\newblock Model-based clustering for social networks.
	\newblock {\em J. R. Statist. Soc. A}, 170:301--354, 2007.
	
	\bibitem{hastie01}
	T.~Hastie, R.~Tibshirani, and J.~Friedman.
	\newblock {\em The Elements of Statistical Learning}.
	\newblock Springer, 2001.
	
	\bibitem{hastings2006community}
	M.~B Hastings.
	\newblock Community detection as an inference problem.
	\newblock {\em Physical Review E}, 74(3):035102, 2006.
	
	\bibitem{Hoff2002}
	P.~D. Hoff, A.~E. Raftery, and M.~S. Handcock.
	\newblock Latent space approaches to social network analysis.
	\newblock {\em Journal of the American Statistical Association}, 97:1090--1098,
	2002.
	
	\bibitem{Hoff2009}
	P.~D. Hoff.
	\newblock Multiplicative latent factor models for description and prediction of
	social networks.
	\newblock {\em Comput. Math. Organ. Theory}, 15(4):261--272, 2009.
	
	\bibitem{Holland83}
	P.~W. Holland, K.~B. Laskey, and S.~Leinhardt.
	\newblock Stochastic blockmodels: first steps.
	\newblock {\em Social Networks}, 5(2):109--137, 1983.
	
	\bibitem{Hu2016}
	J.~Hu, H.~Qin, T.~Yan, and Y.~Zhao.
	\newblock On consistency of model selection for stochastic block models.
	\newblock 2016.
	\newblock arXiv:1611.01238.
	
	\bibitem{Jackson2008}
	M.~O. Jackson.
	\newblock {\em Social and economic networks}.
	\newblock Princeton University Press, Princeton and Oxford, 2008.
	
	\bibitem{Jin2015}
	J.~Jin.
	\newblock Fast network community detection by score.
	\newblock {\em Annals of Statistics}, 43(1):57--89, 2015.
	
	\bibitem{joseph2013impact}
	A.~Joseph and B.~Yu.
	\newblock Impact of regularization on spectral clustering.
	\newblock {\em arXiv:1312.1733}, 2013.
	
	\bibitem{Karlebach08}
	G.~Karlebach and R.~Shamir.
	\newblock Modelling and analysis of gene regulatory networks.
	\newblock {\em Nature Reviews Molecular Cell Biology}, 9, 2008.
	
	\bibitem{Karrer10}
	B.~Karrer and M.~E.~J. Newman.
	\newblock Stochastic blockmodels and community structure in networks.
	\newblock {\em Physical Review E}, 83:016107, 2011.
	
	\bibitem{Kolaczyk2009}
	E.~D. Kolaczyk.
	\newblock {\em Statistical Analysis of Network Data: Methods and Models}.
	\newblock Springer, 2009.
	
	\bibitem{Krzakala13}
	F.~Krzakala, C.~Moore, E.~Mossel, J.~Neeman, A.~Sly,
	L.~Zdeborová, and P.~Zhang.
	\newblock Spectral redemption in clustering sparse networks.
	\newblock {\em Proceedings of the National Academy of Sciences},
	110(52):20935--20940, 2013.
	
	\bibitem{Kumar2004}
	A.~Kumar, Y.~Sabharwal, and S.~Sen.
	\newblock A simple linear time (1+ $\epsilon$) -approximation algorithm for
	k-means clustering in any dimensions.
	\newblock In {\em Proceedings of the 45th Annual IEEE Symposium on Foundations
		of Computer Science}, FOCS '04, pages 454--462, Washington, DC, USA, 2004.
	IEEE Computer Society.
	
	\bibitem{Le2015}
	C.~M. Le and E. Levina.
	\newblock Estimating the number of communities in networks by spectral methods.
	\newblock {\em arXiv:1507.00827}, 2015.
	
	\bibitem{Le16}
	C.~M. Le, E. Levina, and R. Vershynin.
	\newblock Sparse random graphs: regularization and concentration of the
	laplacian.
	\newblock 2015.
	\newblock arXiv:1502.03049v2.
	
	\bibitem{le2017concentration}
	C.~M Le, E. Levina, and R. Vershynin.
	\newblock Concentration and regularization of random graphs.
	\newblock {\em Random Structures \& Algorithms}, 2017.
	
	\bibitem{Lei:2016}
    J.~Lei.
	\newblock A goodness-of-fit test for stochastic block models.
	\newblock {\em Annals of Statistics}, 44(1):401--424, 2016.
	
	\bibitem{Lei&Rinaldo2015}
	J.~Lei and A.~Rinaldo.
	\newblock Consistency of spectral clustering in stochastic block models.
	\newblock {\em Annals of Statistics}, 43(1):215--237, 2015.
	
	\bibitem{Liben2007}
	D.~Liben-Nowell and J.~Kleinberg.
	\newblock The link-prediction problem for social networks.
	\newblock {\em Journal of the American Society for Information Science and
		Technology}, 58(7):1019--1031, 2007.
	
	\bibitem{Massoulie2014}
	L. Massouli{\'e}.
	\newblock Community detection thresholds and the weak ramanujan property.
	\newblock In {\em Proceedings of the 46th Annual ACM Symposium on Theory of
		Computing}, pages 694--703. ACM, 2014.
	
	\bibitem{mossel2013proof}
	E. Mossel, J. Neeman, and A. Sly.
	\newblock A proof of the block model threshold conjecture.
	\newblock {\em arXiv:1311.4115}, 2013.
	
	\bibitem{mossel2015reconstruction}
	E. Mossel, J. Neeman, and A. Sly.
	\newblock Reconstruction and estimation in the planted partition model.
	\newblock {\em Probability Theory and Related Fields}, 162(3-4):431--461, 2015.
	
	\bibitem{mossel2014belief}
	E. Mossel, J. Neeman, A. Sly, et~al.
	\newblock Belief propagation, robust reconstruction and optimal recovery of
	block models.
	\newblock In {\em COLT}, volume~35, pages 356--370, 2014.
	
	\bibitem{mossel2015density}
	E. Mossel and J. Xu.
	\newblock Density evolution in the degree-correlated stochastic block model.
	\newblock {\em arXiv:1509.03281}, 2015.
	
	\bibitem{mossel2016local}
	E. Mossel and J. Xu.
	\newblock Local algorithms for block models with side information.
	\newblock In {\em Proceedings of the 2016 ACM Conference on Innovations in
		Theoretical Computer Science}, pages 71--80. ACM, 2016.
	
	\bibitem{nadakuditi2013spectra}
	R.~Rao Nadakuditi and M.~EJ Newman.
	\newblock Spectra of random graphs with arbitrary expected degrees.
	\newblock {\em Physical Review E}, 87(1):012803, 2013.
	
	\bibitem{Newman2003Review}
	M.~E.~J. Newman.
	\newblock The structure and function of complex networks.
	\newblock {\em SIAM Review}, 45:167--256, 2003.
	
	\bibitem{Newman2004Review}
	M.~E.~J. Newman.
	\newblock Detecting community structure in networks.
	\newblock {\em Eur. Phys. J. B}, 38:321--330, 2004.
	
	\bibitem{Newman2006}
	M.~E.~J. Newman.
	\newblock Finding community structure in networks using the eigenvectors of
	matrices.
	\newblock {\em Phys. Rev. E}, 74(3):036104, Sep 2006.
	
	\bibitem{NewmanPNAS}
	M.~E.~J. Newman.
	\newblock Modularity and community structure in networks.
	\newblock {\em Proc. Natl. Acad. Sci. USA}, 103(23):8577--8582, 2006.
	
	\bibitem{Newman2010}
	M.~E.~J. Newman.
	\newblock {\em Networks: An introduction}.
	\newblock Oxford University Press, 2010.
	
	\bibitem{Newman2016}
	M.E.J. Newman and A.~Clauset.
	\newblock Structure and inference in annotated networks.
	\newblock {\em Nature Communications}, 7, 2016.
	
	\bibitem{Nowicki2001}
	K.~Nowicki and T.~A.~B. Snijders.
	\newblock Estimation and prediction for stochastic blockstructures.
	\newblock {\em Journal of the American Statistical Association},
	96(455):1077--1087, 2001.
	
	\bibitem{Qi_local2014}
	X. Qi, W. Tang, Y. Wu, G. Guo, E. Fuller, and C.-Q.
	Zhang.
	\newblock Optimal local community detection in social networks based on density
	drop of subgraphs.
	\newblock {\em Pattern Recognition Letter}, 36:46--53, 2014.
	
	\bibitem{Qin2013}
	T. Qin and K. Rohe.
	\newblock Regularized spectral clustering under the degree-corrected stochastic
	blockmodel.
	\newblock In {\em Proceedings of the 26th International Conference on Neural
		Information Processing Systems}, NIPS'13, pages 3120--3128, USA, 2013. Curran
	Associates Inc.
	
	\bibitem{Rohe2011}
	K.~Rohe, S.~Chatterjee, and B.~Yu.
	\newblock Spectral clustering and the high-dimensional stochastic block model.
	\newblock {\em Annals of Statistics}, 39(4):1878–--1915, 2011.
	
	\bibitem{saldana2016many}
	D.~Saldana, Y.~Yu, and Y.~Feng.
	\newblock How many communities are there?
	\newblock {\em Journal of Computational and Graphical Statistics}, 2016.
	\newblock To appear.
	
	\bibitem{Sarkar&Bickel2015}
	P. Sarkar and P. Bickel.
	\newblock Role of normalization in spectral clustering for stochastic
	blockmodels.
	\newblock {\em Annals of Statistics}, 43(3):962--990, 2015.
	
	\bibitem{Shi&Malik2000}
	J.~Shi and J.~Malik.
	\newblock Normalized cuts and image segmentation.
	\newblock {\em IEEE Transactions on Pattern Analysis and Machine Intelligence},
	22(8):888--905, 2000.
	
	\bibitem{Snijders&Nowicki1997}
	T.~Snijders and K.~Nowicki.
	\newblock Estimation and prediction for stochastic block-structures for graphs
	with latent block structure.
	\newblock {\em Journal of Classification}, 14:75--100, 1997.
	
	\bibitem{Tzikas2008}
	D.~G. Tzikas, A.~C. Likas, and N.~P. Galatsanos.
	\newblock The variational approximation for bayesian inference.
	\newblock {\em IEEE Signal Processing Magazine}, 25:131--146, 2008.
	
	\bibitem{Laarhoven_local16}
	T. van Laarhoven and E. Marchiori.
	\newblock Local network community detection with continuous optimization of
	conductance and weighted kernel k-means.
	\newblock {\em Journal of Machine Learning Research}, 17:1--28, 2016.
	
	\bibitem{vonLuxburg2007}
	U. von Luxburg.
	\newblock A tutorial on spectral clustering.
	\newblock {\em Statistics and Computing}, 17(4):395--416, 2007.
	
	\bibitem{Bickel:Wang2016}
	Y.X.~Rachel Wang and P.~J. Bickel.
	\newblock Likelihood-based model selection for stochastic block
	modelslikelihood-based model selection for stochastic block models.
	\newblock {\em Annals of Statistics}, 2016.
	\newblock To appear.
	
	\bibitem{Wasserman1994}
	S. Wasserman and K. Faust.
	\newblock {\em Social Network Analysis: Methods and Applications}.
	\newblock Cambridge University Press, 1994.
	
	\bibitem{Wasserman1996}
	S. Wasserman and P. Pattison.
	\newblock Logit models and logistic regressions for social networks. {I}. {A}n
	introduction to {M}arkov graphs and {$p^*$}.
	\newblock {\em Psychometrika}, 61(3):401--425, 1996.
	
	\bibitem{Wei&Cheng1989}
	Y.-C. Wei and C.-K. Cheng.
	\newblock Toward efficient hierarchical designs by ratio cut partitioning.
	\newblock In {\em Proceedings of the IEEE International Conference on Computer
		Aided Design}, pages 298--301, 1989.
	
	\bibitem{Wu_local15}
	Y. Wu, R. Jin, J. Li, and X. Zhang.
	\newblock Robust local community detection: On free rider effect and its
	elimination.
	\newblock {\em Proceedings of the VLDB Endowment}, 8(7):798--809, 2015.
	
	\bibitem{Xu2012}
	Z. Xu, Y. Ke, Y.~Wang, H. Cheng, and J. Cheng.
	\newblock A model-based approach to attributed graph clustering.
	\newblock In {\em Proceedings of the 2012 ACM SIGMOD International Conference
		on Management of Data}, SIGMOD '12, pages 505--516, New York, NY, USA, 2012.
	ACM.
	
	\bibitem{Sarkar16}
	B. Yan and P. Sarkar.
	\newblock Convex relaxation for community detection with covariates.
	\newblock 2016.
	\newblock arXiv:1607.02675v3.
	
	\bibitem{Yang13}
	J.~Yang, J.~McAuley, and J.~Leskovec.
	\newblock {Community Detection in Networks with Node Attributes}.
	\newblock In {\em IEEE International Conference On Data Mining (ICDM)}, 2013.
	
	\bibitem{yedidia2003understanding}
	J.~S Yedidia, W.~T Freeman, and Y. Weiss.
	\newblock Understanding belief propagation and its generalizations.
	\newblock {\em Exploring artificial intelligence in the new millennium},
	8:236--239, 2003.
	
	\bibitem{zhang2016minimax}
	A.~Y Zhang, H.~H Zhou, et~al.
	\newblock Minimax rates of community detection in stochastic block models.
	\newblock {\em Annals of Statistics}, 44(5):2252--2280, 2016.
	
	\bibitem{zhang2014}
	X. Zhang, R.~R. Nadakuditi, and M.~EJ Newman.
	\newblock Spectra of random graphs with community structure and arbitrary
	degrees.
	\newblock {\em Physical Review E}, 89(4):042816, 2014.
	
	\bibitem{Zhang2016}
	Y. Zhang, E. Levina, and J.~Zhu.
	\newblock Community detection in networks with node features.
	\newblock {\em Electronic Journal of Statistics}, 10:3153–3178, 2016.
	
	\bibitem{Zhao.et.al.2011}
	Y.~Zhao, E.~Levina, and J.~Zhu.
	\newblock Community extraction for social networks.
	\newblock {\em Proc. Natl. Acad. Sci. USA}, 108(18):7321--7326, 2011.
	
	\bibitem{Zhaoetal2012}
	Y.~Zhao, E.~Levina, and J.~Zhu.
	\newblock Consistency of community detection in networks under degree-corrected
	stochastic block models.
	\newblock {\em Annals of Statistics}, 40(4):2266--2292, 2012.
	
\end{thebibliography}
	\end{document}